\documentclass[a4paper,10pt,journal]{IEEEtran}
\usepackage{amsmath,amsfonts}
\usepackage{algorithmic}
\usepackage{algorithm}
\usepackage{balance}
\usepackage{array}
\usepackage[caption=false,font=normalsize,labelfont=sf,textfont=sf]{subfig}
\usepackage{textcomp}
\usepackage{stfloats}
\usepackage{url}
\usepackage{verbatim}
\usepackage{graphicx}
\usepackage{cite}
\author{Faisal Zaman$^1$,   Ouns Bouachir$^2$, Moayad Aloqaily$^3$, and Ismaeel Al Ridhawi$^4$\vspace{5px} \\
{$^1$University of Montreal, Canada}\\
 {$^2$Zayed University, UAE}\\
{$^3$United Arab Emirates University (UAEU), Al Ain, UAE.}\\
{$^4$Kuwait College of Science and Technology, Kuwait}\\
  Emails: {faisal.zaman@umontreal.ca; ouns.bouachir@zu.ac.ae; maloqaily@uaeu.ac.ae; i.alridhawi@kcst.edu.kw
    }
 
 }
\begin{document}

\title{Generative AI for Networking}

%\author{IEEE Publication Technology,~\IEEEmembership{Staff,~IEEE,}
        % <-this % stops a space
% \thanks{This paper was produced by the IEEE Publication Technology Group. They are in Piscataway, NJ.}% <-this % stops a space
% \thanks{Manuscript received April 19, 2021; revised August 16, 2021.}}
%}
% The paper headers
% \markboth{Journal of \LaTeX\ Class Files,~Vol.~14, No.~8, August~2021}%
% {Shell \MakeLowercase{\textit{et al.}}: A Sample Article Using IEEEtran.cls for IEEE Journals}

% \IEEEpubid{0000--0000/00\$00.00~\copyright~2021 IEEE}
% Remember, if you use this you must call \IEEEpubidadjcol in the second
% column for its text to clear the IEEEpubid mark.

\maketitle

\begin{abstract}
Generative Artificial Intelligence (GenAI) and Large Language Models (LLMs) are revolutionizing network management systems, paving the way towards fully autonomous and self-optimizing communication systems. These models enable networks to address complex decision-making tasks across both short-term operational scenarios and long-term strategic planning. Through natural language understanding, LLMs can analyze customer inquiries, predict network congestion patterns, and automate troubleshooting processes, leading to more efficient customer support and network maintenance. GenAI can optimize content delivery by generating personalized recommendations, improving user engagement, and dynamically adjusting network resources based on real-time demands, ultimately enhancing overall performance and user experience in telecommunication services. In this paper, we discuss the pivotal role of GenAI in advancing network performance and achieving the ultimate objective of self-adaptive networks. Moreover, we present a use case that leverages the self-attention mechanism of transformers to perform long-term traffic prediction. Harnessing these cutting-edge technologies demonstrates the transformative power of LLM and GenAI in revolutionizing telecommunication networks, elevating resilience and adaptability to unprecedented levels.
%The introduction of Generative Artificial Intelligence (GenAI) and Large Language Models (LLMs) enables a new era in network management capable of addressing complex decision-making challenges essential for both long- and short-term network optimization. LLMs can enhance telecommunication performance by optimizing various aspects of network management, customer service, and content delivery. Through natural language understanding, these models can analyze customer inquiries, predict network congestion patterns, and automate troubleshooting processes, leading to more efficient customer support and network maintenance. In addition, they can optimize content delivery by generating personalized recommendations, improving user engagement, and dynamically adjusting network resources based on real-time demands, ultimately enhancing overall performance and user experience in telecommunication services. In this paper, we discuss the pivotal role of GenAI in advancing network performance and achieving the ultimate objective of self-adaptive networks. In addition, we present a use case that leverages the self-attention mechanism of transformers to perform long-term traffic prediction. Harnessing these cutting-edge technologies demonstrates the transformative power of LLM and GenAI in revolutionizing telecommunication networks, elevating resilience and adaptability to unprecedented levels.
 \end{abstract}

\section{Introduction}
The rapid emergence of advanced technologies has increased the complexity of modern network management. %The advent of new technologies has also exponentially increased the complexity of managing networks. 
Various software tools are used in improving network management and optimization. The most commonly used technique is to leverage descriptive models. Descriptive-based Artificial Intelligence (AI) models have labels and can assist in making an informed decision about providing information on outages, channel capacity enhancement, and classification of new traffic types. But as the complexity and heterogeneity of managing self-organizing networks increases, a need to explore other techniques and approaches is required. Generative AI, including tools like ChatGPT and Gemini, has undeniably reshaped the technology landscape, unlocking transformational use cases, such as creating original content, generating code, and expediting customer service. The scope of this technology continues to expand on a daily basis. %}

%Organizations that harness this transformative technology successfully, will be differentiated in the marketplace and assume leadership in the future.}

Generative models are focused on creating new data or simulating complex systems to improve decision making and optimize network performance in telecommunications. Several proposals have been made to take advantage of this generative capability in order to improve network performance. Most of these approaches focus on how GenAI technology can enhance the user experience or offer a high-level overview of various network solutions. 

The authors in \cite{survey_gen_ai} provide an overview of different GenAI solutions such as the generation of text, audio, images, and video, in addition to the interconversion between them. With that said, there is still a lack of discussion on the use of GenAI in the context of networking. In \cite{survey_1}, \cite{survey_2}, and \cite{survey_3}, the use of LLMs is discussed primarily from a conceptual perspective, emphasizing theoretical viewpoints while providing limited analysis of existing practical solutions. In \cite{survey_3}, the authors discuss various aspects of the communication channel from a wireless perspective, where GenAI can be applied, such as the receiver (Rx), transmitter (Tx) and communication channels. They also provide a use-case for GenAI in Tx and Rx to improve Signal-to-Noise Ratio (SNR). In \cite{survey_4}, the use of LLMs is outlined along with a demonstration of a simple use case. Various applications of LLMs for networking verticals are discussed, and key enabling technologies such as prompt engineering are highlighted. ChatNet is proposed as a system that translates prompts into network planning designs. The focus is predominantly on language models, with evaluations conducted using GPT-4 to illustrate network planning, assessing the ML model rather than the results' outcomes. Moreover, GANs are utilized in the framework to estimate generative models through an adversarial process involving the simultaneous training of a discriminator and a generator to identify anomalies in the data. 
The work in \cite{survey_5} discusses several GAN architectures used to determine anomalies in the data. The paper provides an overview of how GANs can be used for anomaly detection and does not focus on telecommunication aspects.

To the best of our knowledge, there is a lack of discussion on the use of GenAI with LLM to improve network management in the literature along with use-cases. %Furthermore, use-cases are not presented no one of the existing literature discusses the available solutions using generative AI with large language models and different use cases to improve networking. 
In this paper, we propose a long-term traffic prediction use-case for end-to-end network slice handling in Beyobd 5G (B5G) networks. In summary, the contributions of this paper are as follows:
\begin{enumerate}
    \item Provide a survey of different proposals that use generative AI to provide improved and enhanced end-to-end telecommunication networks. 
    \item Discuss different avenues and gaps that need to be investigated with GenAI.
    \item Present a use-case for traffic prediction and application of optimal policies for end-to-end network slices using transformers.
    % \item Obtained results are compared with the traditional static policy approach. 
\end{enumerate}

\section{Generative AI to Improve End-End Networking}
\subsection{What is Generative AI ?}
\begin{figure}[h]
    \centering
    \includegraphics[width=\columnwidth]{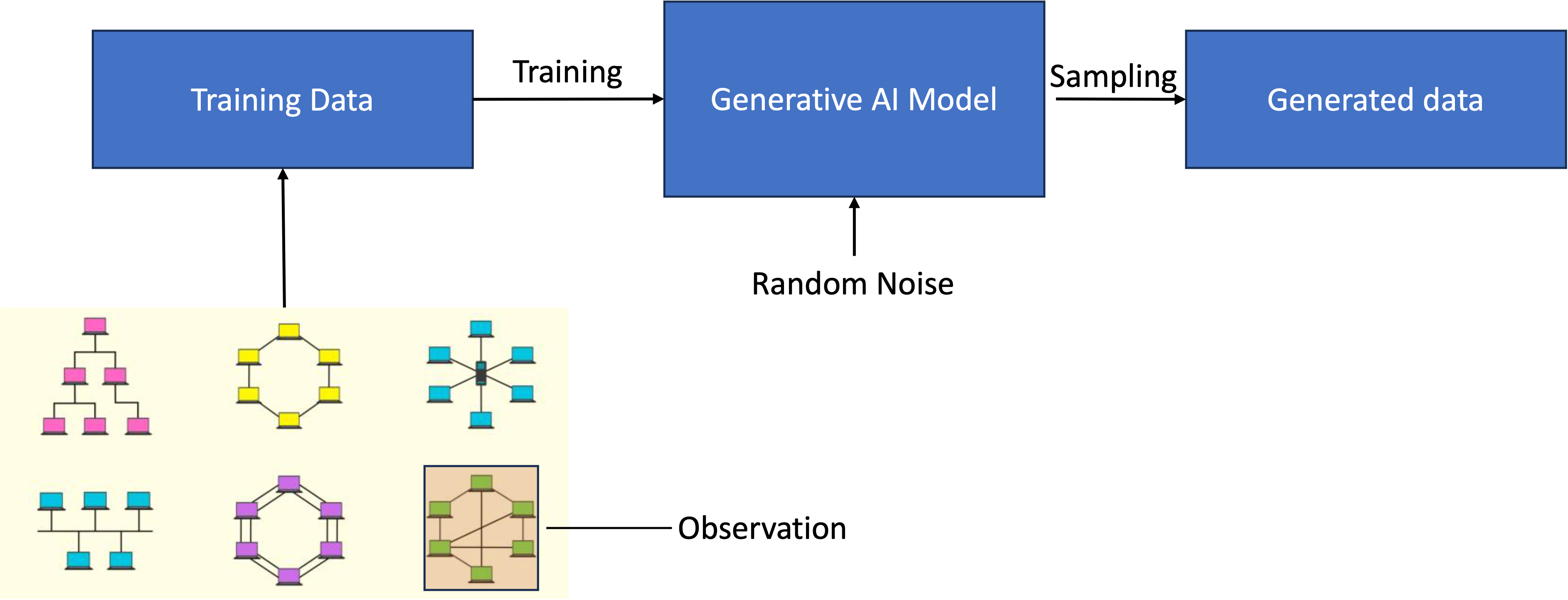}
    \caption{Overview of GenAI process.}
    \label{fig:process_gen}
\end{figure}

Generative models are a class of neural networks that are capable of generating new data from the pretrained data. The overall process of the GenAI model is shown in Figure \ref{fig:process_gen}. Each observation consists of many features (i.e., for an image generation, the features are usually the individual pixel values). It is our goal to build a model that can generate new sets of features that look as if they have been created using the same rules as the original data. Conceptually, for image generation this is an incredibly difficult task, considering the vast number of ways that individual pixel values can be assigned and the relatively tiny number of such arrangements that constitute an image of the entity we are trying to simulate. Sampling refers to the process of generating new data samples from a learned probability distribution. This distribution represents the underlying patterns and structure of the training data on which the generative model has been trained. Sampling is a fundamental aspect of generative models because it allows them to create new synthetic data points that are similar to the data they were trained on. By drawing samples from the learned distribution, the generative model can produce outputs that exhibit characteristics, patterns, and statistical properties similar to the original data.

The specific method used for sampling can vary depending on the type of generative model used. For example, in variational autoencoders (VAEs), sampling involves drawing samples from the latent space distribution. This is typically done by sampling from a simple distribution (e.g., Gaussian) and then transforming these samples through the decoder network. In generative adversarial networks (GANs), sampling involves feeding random noise into the generator network and generating fake data samples. Generative AI can be classified in different ways, such as categorizing according to the application or based on the underlying model used. Based on different types of model, we can broadly classify them into five main types, namely, \textit{i)} variational autoencoders, \textit{ii)} generative adversarial networks, \textit{iii)} autoregressive models, \textit{iv)} flow-based models, and \textit{v)} energy-based models. GenAI models that focus explicitly on determining the probability density function are categorized as explicit models. Whereas ones such as GAN are implicit and rely on an approximation of the latent space. 

\subsection{Generative AI vs Descriptive AI}
Generative AI is a branch of AI that can create new content or output based on patterns it has learned from existing data. This type of AI can generate text, images, and music, to name a few, and mimics the style of the original data it was trained on. This means that generative AI can produce original and creative content that is not directly copied from existing data. Most of the large models such as LLMs including ChatGPT models fall under the category of Generative AI.  Mathematically, the probability can be described as as P(x), where x is the probability of an event to occur. In contrast, a descriptive AI model can analyze and summarize existing data or provide insights based on patterns it has identified in the data. This type of AI does not create new content, but rather describes and explains the existing data on which it has been trained. Descriptive AI is commonly used for tasks such as data analysis, trend prediction, and decision-making based on data patterns. The probability can be described as P(x $|$ y).
% \begin{figure}
%     \centering
%     \includegraphics[width=\columnwidth]{Images/overview.png}
%     \caption{Overview of GenAI classes}
%     \label{fig:LevelGenAI}
% \end{figure}

\subsection{Improving Communication using Generative AI Models}
\begin{figure}[h]
    \centering
    \includegraphics[width=\columnwidth]{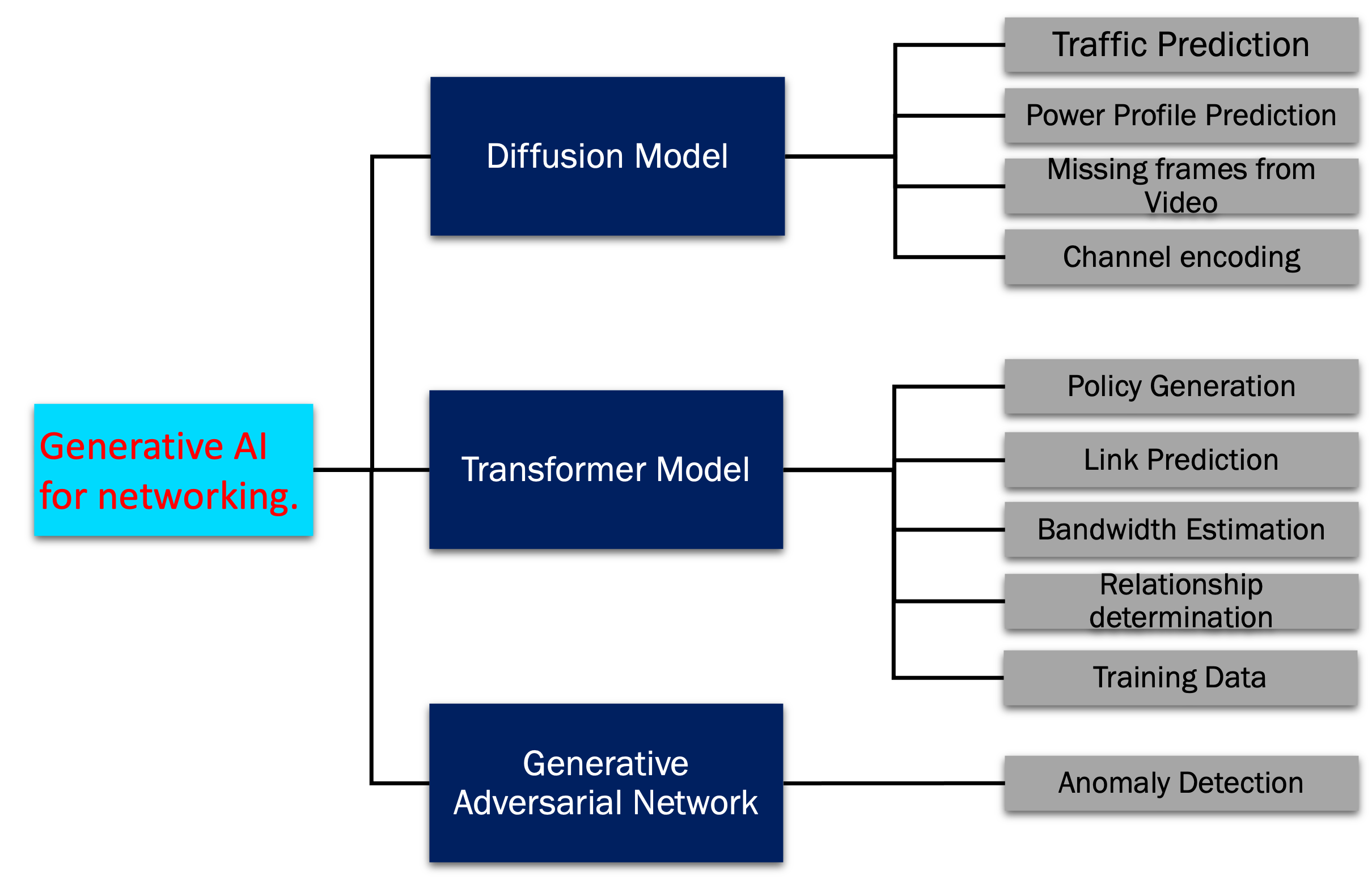}
    \caption{Different Networking AI applications for Networking.}
    \label{fig:TaxonomyGenAI}
\end{figure}

The application of GenAI for improving communications can be broadly classified as shown in Fig. \ref{fig:TaxonomyGenAI}. The three categories can be used to generally categorize the application of GenAI; however, they are not exclusive. 

\subsubsection{Transformer Model}
These belong to a class of deep neural networks. The primary challenge addressed by the transformer model is the parallelization of tasks through the use of the attention layer, which was not possible with the existing models for sequential data \cite{attention_paper}. The key components of this model are: Attention layer, Encoder-Decoder, Positional encoder, Multi-head attention, and Feed-forward network. Self-attention allows the model to weigh the importance of different words in a sequence, enabling it to capture contextual relationships effectively. The feed-forward neural networks provide non-linear transformations of the self-attention outputs, facilitating more complex feature representations. Additionally, positional encoding is used to incorporate sequence-order information. The model is structured into multiple layers of these blocks, with each layer iteratively refining the representation of the input sequence. Several use cases exist in the networking domain:
\begin{enumerate}
    \item \textbf{Policy Generation}:
    Policy can be defined as network rules or directions that dictate how traffic flows through network links. OpenFlow rules, P4 rules, and other forwarding rules, such as multi-protocol label switching (MPLS) and segment routing, decide how network traffic will be routed. The network configuration and modeling rules can be cumbersome, mainly during the congestion interval. To anticipate the network traffic increase and to define network policy to handle different service providers to support a self-optimizing network from an end-to-end network perspective is crucial to support current network demands \cite{policy_transform_1}. In \cite{policy_transform_2}, the authors demonstrated the automation of intent-based decomposition and execution of network policies. They compared the proposal with the few-shot capability of LLMs. To solve the problem in hand, they proposed a pipeline that progressively decomposes intents by generating the required actions using policy-based abstraction. This assisted in the automation of policy execution by creating a closed control loop for the intent of deployment. Talha et al. \cite{policy_transform_3} proposed such an automated approach to model policy.
    \item \textbf{Traffic Prediction}:
    To circumvent the limitation often associated with traffic prediction using the time series methods such as Recurrent Neural Network (RNN) and Long Short-Term Memory (LSTM), LLM and transformers have been utilized to bring better relationships between higher degrees of parameters. In \cite{traffic_predic_1}, the authors proposed a solution called the heterogeneous and temporal model based on the transformer (H-Trans) which aims to better integrate neighbor information to capture the structure of the network. The system is compared with the baseline for the link prediction task on three real datasets. The idea is to predict network links in a complicated, densely connected network. This can be widely used for wireless networks. The proposed model does not discuss the networks related to telecommunications and provides a general link prediction mechanism for heterogeneous graphs. Transformers can be used to predict traffic on a given link with the help of different inputs. The authors in \cite{traffic_predic_2} proposed Autoformer, which is a transformer-based traffic prediction method in O-RAN. The predicted traffic is used to control the sleep cycle of certain energy-intensive O-RAN applications, such as traffic steering, which typically runs for the entire duration. The scope of the paper is limited by the prediction of traffic in a specific domain. The authors focused on short-term traffic prediction and do not discuss the scenario in case of prediction failure. 
    \item \textbf{Synthetic Data}:
    One of the challenges that may arise when integrating AI with telecommunications is the availability of data. With the use of LLM models, few proposals have been made around generating synthetic network data that can be used to train the network models. The work in \cite{item_8} trained GPT-3 to generate ICMP and DNS packets. The generated packets were converted to flow using Python scripts. Since there is limited availability of the training data, this can help in simulating and testing networks. Only two types of network packets were trained. Training data for generating different types of network packets was limited. 

    % \item \textbf{Relationship Determination}
\end{enumerate}
\subsubsection{Diffusion Model}
Diffusion models are a class of generative models that operate by iteratively refining a probability distribution over the data space. They were introduced to address the challenge of generating high-quality samples, especially in the domain of image generation. The key idea behind diffusion models is to model the data generation process as a series of diffusion steps, where noise is progressively added to the data until the probability distribution of the true data is achieved.
The underlying concepts of the diffusion model can be leveraged to solve some of the challenges in communication, which are:
\begin{enumerate}
    \item \textbf{Traffic Prediction}: Traffic prediction has been one of the most investigated topics. Several ML approaches can be used to predict traffic, but one of the limitations of traditional supervised learning is the non-existence of an attention mechanism. The authors in \cite{diff_traffic_1} have used the diffusion model to predict traffic. The authors leverage the diffusion model to determine the topological relations of the links and to predict the next step of congestion. To include the nuances of the network graph, the authors included traffic attributes as part of representing the directed network graph. The model used is the diffusion-based Convolution Neural Network (CNN). The results show that the inclusion of different graph relationships can drastically improve the traffic prediction model. 
    % \item \textbf{Missing Frames} to generate video frames during the transmission of high-definition videos in a lossy network.
    \item \textbf{Channel Encoding and Compression}: Semantic communication refers to the exchange of meaning between individuals or entities through various forms of language or symbols. In telecommunications, it is used to reduce redundant information transmission and to improve channel utilization. Several proposals have leveraged AI to improve existing semantic communication \cite{item_11}.  C. Dong et al. \cite{item_10} leveraged generative AI to determine semantic relationships.  The semantic communication system is designed based on the Shannon-based physical layer, and the AI models are integrated into the transmitter and receiver. Model propagation is the main feature of the proposed solution. The authors also proposed metrics called semantic service quality to evaluate the performance and accuracy of semantic-based communication. The solution was tested with image compression and transmission. 
    % \item \textbf{Power Profile Prediction} This is more re
\end{enumerate}

\subsubsection{Generative Adversarial Network}
GANs consist of two key architecture blocks: the generator and the discriminator. The generator generates synthetic data samples by mapping random noise vectors to output data representations, in an attempt to mimic the distribution of real data. The discriminator, on the other hand, is trained to distinguish between real and fake data samples. It provides feedback to the generator by assessing the realism of its generated samples. This adversarial training process encourages the generator to produce increasingly realistic samples, while the discriminator becomes more adept at distinguishing between real and fake data. The interaction between these two components leads to the convergence of the generator towards generating highly realistic data samples. Random noise refers to a set of randomly generated input values that are fed into the generator network. This noise typically follows a Gaussian or uniform distribution. The purpose of this noise is to provide the generator with a source of randomness, allowing it to produce various outputs during the generation process.
One of the most commonly discussed applications of GAN in telecommunication is its use for anomaly detection. The authors in \cite{item_9}, proposed a GAN-based solution for anomaly detection in binary fashion in networks. The proposed solution consists of a property scaling module and a GAN-based detection module. The scaling module selects features to detect anomalies effectively in a minority of cases. The detection module detects to cope with class imbalances in intrusion detection. They also use GAN to generate new data.

%\subsubsection{Summary and future direction}
Various LLM-based models have been employed in numerous studies to enhance network performance. There are still several areas that need to be investigated in GenAI-enabled telecommunication. Some of the areas are: %that require further investigation based on various ideas and a recent survey about LLM are as follows.  
\begin{enumerate}
    \item \textbf{GenAI for Optical Communications}: Our existing network is largely based on optical communication. More than 90\% of global communication occurs through fiber networks. Focusing on topics such as creating the best channel strategies, optimizing regions, and refining location optimization for greenfield deployment could be beneficial.  
    \item \textbf{Relationship Determination}: One of the challenges of machine learning is the relationship determination of models with a huge set of influencing parameters. For example, capricious traffic changes can be due to several factors in the area. To accurately predict this using traditional models can be computationally impractical, but the use of LLMs has proven otherwise. Using an attention layer with a transformer is an example of how a sequentially related task can be converted into parallel tasks. Similar approaches can be used to create relationships between different events and parallelize the workflow.
    \item \textbf{Confluence of different GenAI Approaches}: GenAI for telecommunications is still in its infancy and focuses on using a single type of model for solving specific solutions. The ultimate goal is to create a self-optimizing network. To achieve such a goal, it is required to use more than one type of model and to work symbiotically. 
\end{enumerate}
%The above is a nonexhaustive list that broadly covers a few topics that are not considered in the existing work. In the next section, we explore the use case of applying a transformer for traffic prediction and policy generation for OpenFlow in an end-to-end network.  
% \subsection{Role of AI}
% \section{Generative AI for self Adaptive Networks}

\section{Usescase: Policy Generation for End-to-End Network Slicing}
In the context of a network service provider that oversees multiple network slices, each endowed with constrained resources, dynamic management of these resources in network slices is imperative. This entails the scaling of network slices. However, for effective resource allocation across various network slices, the provider must forecast demand within each slice. Long-term forecasting is extremely important, as it provides enough time for service providers to make policy changes and infrastructure upgrades. Many of the existing solutions such as Informer\cite{informer_1} and other averaging methods such as root mean square do not capture the peaks or do not consider different parameters to accurately predict long-term forecasts.  Our proposal primarily addresses the challenge of predicting long-term network traffic. %In the following subsection, we describe the key functional blocks of our proposed system model. 

\begin{figure}[t]
    \centering
    \includegraphics[width=\columnwidth]{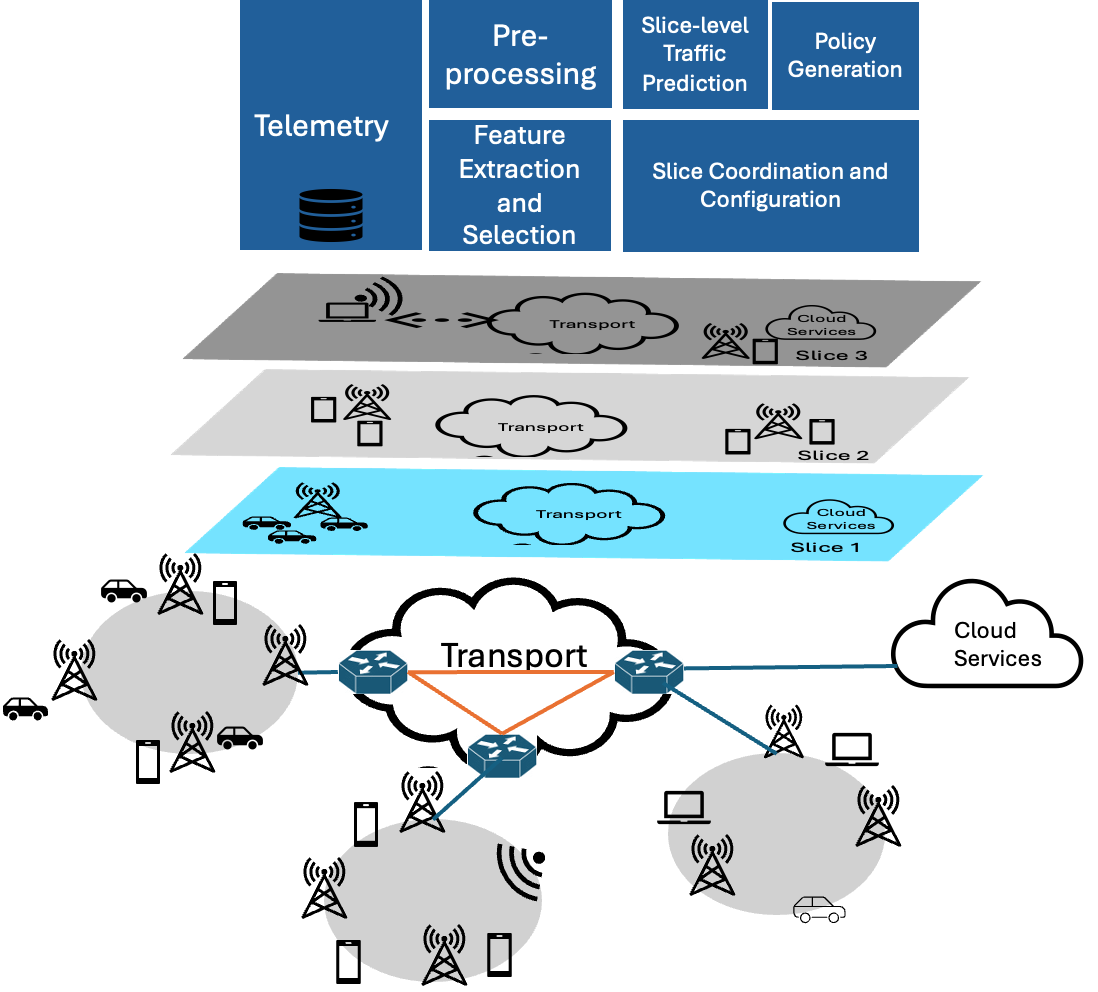}
    \caption{Functional Block Diagram of System Model and Network Slice Management.}
    \label{fig:systemModel}
\end{figure}

\subsection{System Model}
The proposed system model for training and generating policies to scale different slices is shown in Figure \ref{fig:systemModel}. The system comprises of the following:
\begin{itemize}
    \item \textbf{Telemetry: } It is responsible for collecting metrics from the system. Different databases are available to collect and store data. We leverage the Prometheus database to manage our time-series metrics. Prometheus uses its own time-series database written in Go, which is optimized for high performance and efficient storage of time-series data. This repository serves as the central repository wherein all metrics garnered by Prometheus are deposited and subsequently retrieved for analysis and interpretation. The input to telemetry can be captured from all the network elements that are part of the service provider that owns all different network slices. The different elements of designing telemetry solutions and design considerations on the types of data collection utilized, the granularity, and other aspects are beyond the scope of this paper. We assume that the overhead of collecting and managing data from the network is negligible. The data from the telemetry is fed to the preprocessing and feature extraction. 
    \item \textbf{Preprocessing: } This step plays a key role in ensuring that the data is normalized and can be utilized by the AI engine. Several steps must be performed. The key preprocessings considered in our model are data cleaning, temporal alignment, normalization, and aggregation.
    \item \textbf{Feature selection: } Depending on the problem considered, we can select different features to get the best results. In most cases, feature selection is part of preprocessing. In our scenario, we consider the rate of packet flow as the feature to predict the traffic in relation to time.
    \item \textbf{Slice-level Traffic Prediction: } This represents our AI models in consideration. We utilize Autoformers to predict long-term traffic. As pointed out, the existing models have a limitation when predicting for long-term \cite{traffic_predic_2}. The authors in \cite{autoformers_1} prove that transformer-based models show promise in capturing long-term dependencies, they face computational constraints and struggle with intricate temporal patterns. To address these issues, the authors propose Autoformer. This architecture combines decomposition techniques with an Auto-Correlation mechanism. This enables the progressive decomposition of time series and series-level information aggregation based on periodicity, enhancing both accuracy and computational efficiency. Hence, we leverage Autoformers in our work for long-term traffic prediction. The other reason for using Autoformers as apposed to other GenAI based approach for traffic prediction are described in the table \ref{tab:autoformer}.  The predicted traffic output should be utilized to reconfigure existing network slices to ensure that the right number of slices is available to support the upcoming traffic change.  
\begin{table}[]
    \caption{Autoformer vs Other Transformer based GenAI for traffic prediction}
    \centering
    \begin{tabular}
    {|p{0.45\columnwidth}|p{0.45\columnwidth}|}
    \hline
         Autoformer & Other GenAI (e.g., Informer, LogTrans)   \\
         \hline
         \hline
         Auto-correlation Layer - responsible for determining dependencies and aggregate information at the sub-series level based on periodicity  & Uses Self-Attention layer which discovers dependencies and aggregates information at the point-wise level. \\
         \hline
         Uses deep Decomposition block as an inner block to separate intricate temporal patterns &
         Has found to be unreliable for finding long-term temporal patterns such a seasons and trends leading to entangled temporal patterns. \\
         \hline
         Has a $O(L Log L)$ where L is the length of time-series & Can be from $O(L^2)$ to $O(L Log L)$ depending on different layers an use cases. Hence this requires more than one type of transformer model to predict traffic. \\
         \hline
         Better suited for capturing seasonality of complex real-world using hourly based data & cannot handle seasonality prediction \\
         \hline
         % Deep Decomposition Block & Yes & No \\
         
    \end{tabular}

    \label{tab:autoformer}
\end{table}

    \item \textbf{Policy Generation: } To create policies that can be easily translated to network setup, we employ various LLM models to handle traffic for network devices. The key benefit of using LLM is the accuracy of the policy generated for the heterogeneous network. 
    \item \textbf{Slice Coordination and Configuration: } This module coordinates with different network agents in the right order to ensure that the policy gets applied to the network elements. This also includes coordinating with the infrastructure service provider in parts that are not owned by the service provider.  
\end{itemize}

%In the following section, we discuss the Implementation aspect of the system model. 

\begin{figure}[t]
    \centering
    \includegraphics[width=0.4\columnwidth]{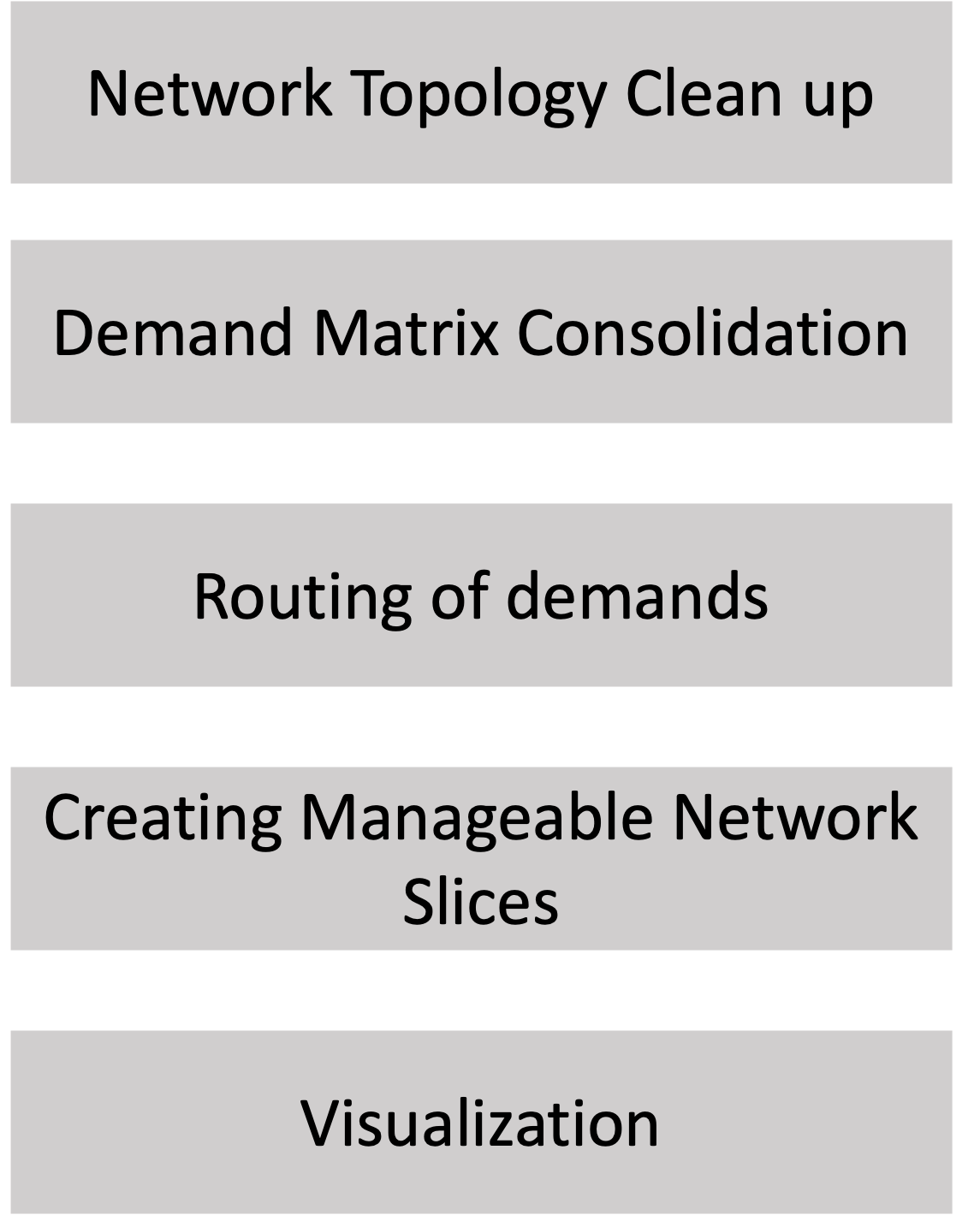}
    \caption{Major Steps Used for Model Preprocessing.}
    \label{fig:model_steps}
\end{figure}

\subsection{Implementation}
As a proof of concept, we implemented an Autoformer-based traffic prediction method. The results are compared with the Informer-based approach.  The traffic data is generated based on the demand matrix available on the SDNLib. The simulation and other preprocessing are done using Python, leveraging libraries such as PyTorch, Matplotlib, and Pandas. The data is divided using the standard approach of 60:20:20 for training, validation, and testing. %Different steps involved in preprocessing and training are described in the following section.
To generate network data for simulation purposes we performed several key steps. These steps are depicted in Figure \ref{fig:model_steps}.

\begin{figure}[t]
    \centering
    \includegraphics[width=0.8\columnwidth]{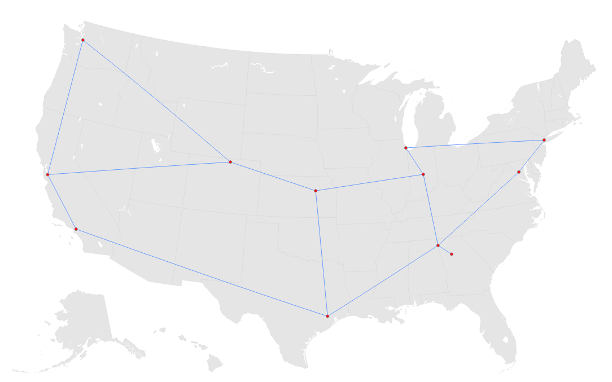}
    \caption{Albeini Network used for Analysis.}
    \label{fig:network}
\end{figure}
\begin{figure}[b]
    \centering
    \includegraphics[width=0.8\columnwidth]{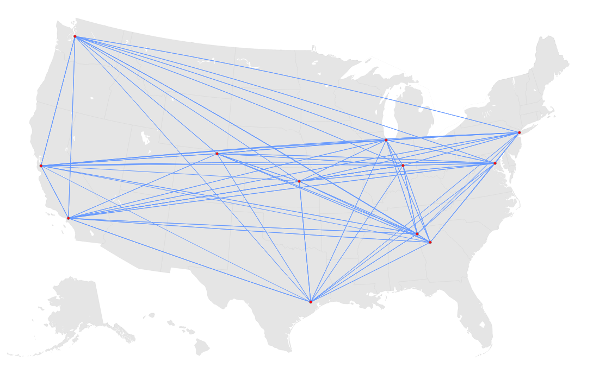}
    \caption{Traffic Demand Matrix.}
    \label{fig:demand}
\end{figure}

%\subsubsection{Preprocessing steps}
\begin{enumerate}
    \item \textbf{Network Topology Selection}: We used the Albenin topology retrieved from the archive \cite{sdnlib} for network topology. It records network demand every five minutes along with the latitude and longitude positions of the network. The network connectivity is shown in Figure \ref{fig:network}. The topology was cleaned and the connectivity and relevant traffic data were extracted from the archive. 
    \item \textbf{Demand Matrix Consolidation}: The data available was for every five minutes for six months. For a proof-of-concept, we consolidated this data and took the maximum of each day for each demand. The demand matrix is shown in Figure \ref{fig:demand}. 
    \item \textbf{Routing of Demand}: To map the topology connectivity with the routing of demands, we used the Dijkstra algorithm and consolidated common path traffic to create several manageable network slices. Once we have the view of these slices, we added the visualization to understand the routing and demand requirement for each slice. 
    \item \textbf{Network Slices}: The common routes are consolidated and different network slices were formed. Each slice can serve multiple services of similar Quality of Service (QoS) classes. In our context, we are not focused on the type of network slices, but in the future, we will be including this parameter to improve the prediction algorithm.
\end{enumerate}
% The key functional steps performed for the data 

\subsection{Limitations}
Although Autoformer demonstrates strong potential for network traffic prediction, several key practical considerations must be addressed to ensure its effective deployment in real-world networks especially in wireless B5G networks. For instance, in such environments, traffic dynamics are influenced not only by data flow but also by interference, which can significantly distort predictions and lead to false traffic estimations.

Efficient data collection is another critical factor. Long-term time series prediction requires high-quality data across multiple network components, including traffic type, bandwidth utilization, interference levels, user mobility, and sampling frequency. Each of these parameters directly impacts data quality and, consequently, the accuracy of the predictive model.

Hyperparameter optimization remains a major challenge. Parameters such as the autocorrelation factor and input sequence length must be carefully tuned, often through extensive sensitivity analyses. While RL techniques can be employed to automate this process, their integration introduces additional complexity and computational overhead.

Furthermore, Autoformer models exhibit limited capability in data augmentation. To address this, complementary GenAI techniques particularly those based on LLMs can be leveraged to synthesize realistic training data, thereby enhancing the robustness and generalization of Autoformer based predictors.

\subsection{Results and Discussion}
\begin{figure}
    \centering
    \includegraphics[width=0.75\columnwidth]{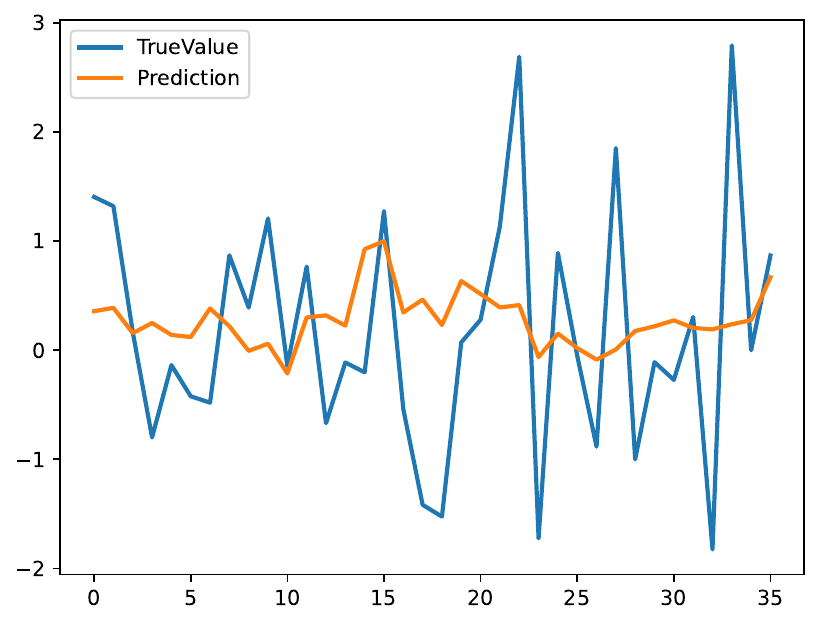}
    \caption{Prediction using Autoformers Model on Network traffic for the short duration 36-time steps}
    \label{fig:ShortTermAuto}
\end{figure}

\begin{figure}
    \centering
    \includegraphics[width=0.75\columnwidth]{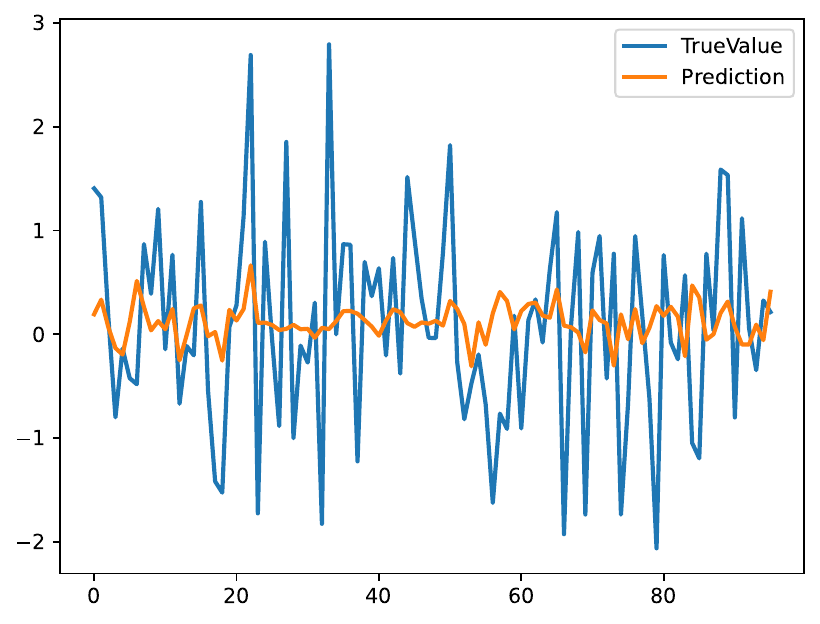}
    \caption{Prediction using Autoformers Model on Network traffic for the short duration 96 time steps}
    \label{fig:LongTermAuto}
\end{figure}

\begin{figure}
    \centering
    \includegraphics[width=0.75\columnwidth]{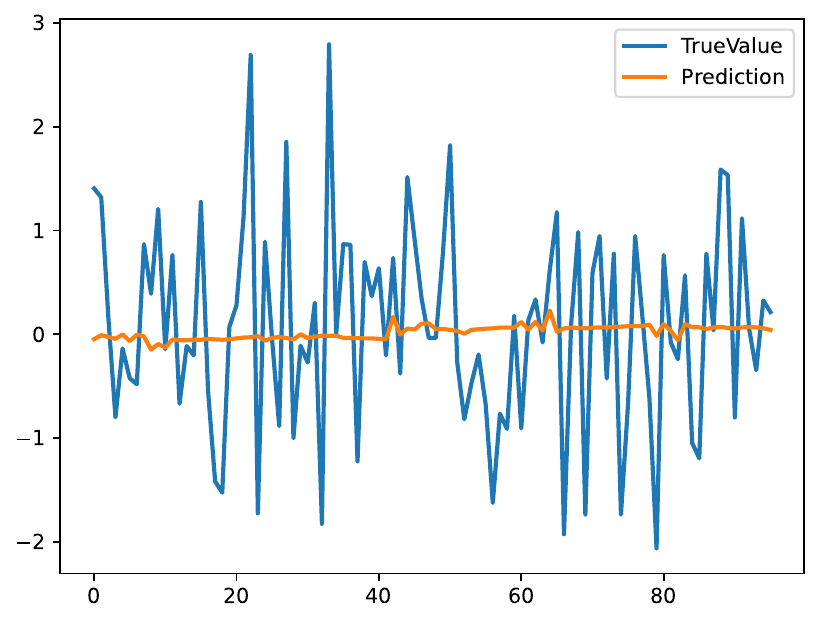}
    \caption{Prediction of Informer Model on Network traffic for short duration 36-time steps}
    \label{fig:informer}
\end{figure}

The results obtained from Autoformer\cite{autoformers_1} are shown in Figures \ref{fig:ShortTermAuto} and \ref{fig:LongTermAuto}. The x-axis refers to the Standard Scalar value obtained by removing mean and scaling to variance. This helps in normalizing the data. The y-axis refers to the time steps. Figure \ref{fig:ShortTermAuto} shows the performance of Autoformer in predicting a shorter duration of about 30 time steps. Whereas Figure \ref{fig:LongTermAuto} shows the long-term view of the prediction. The Autoformer is optimized for long-term prediction, and hence we see that the prediction is relatively flat for short-term results. Whereas for long-term predictions, it follows the variations in traffic. It is clear from the results that the predictor follows the peaks and the overall traffic pattern. % are followed by the predictor.
For the Informers, shown in Figure \ref{fig:informer}, we see that the smoothing algorithm removed the peaks and flattens the traffic pattern giving a wrong signal. Even though the algorithm is not predicting at the highest accuracy, we believe that by including necessary parameters in the input, it can be proved that the long-term prediction can easily be achieved using Autoformers. 

\section{Conclusion}

This paper provided a comprehensive survey of various proposals, leveraging generative AI (GenAI) to enhance communication networks through end-to-end network optimization. It highlighted key avenues and identifies gaps that require further investigation to fully realize the potential of GenAI in the networking sector. Furthermore, the paper presented a practical use case involving traffic prediction and the application of optimal policies for end-to-end network slicing using a transformer, demonstrating the practical utility of GenAI in improving network efficiency and performance. In the future, we will evaluate the proposed policy generation mechanism and different network slice parameters. We will also integrate the solution with an emulated testbed using Mininet. We also plan to focus on extending the experimental scope of the proposed framework by evaluating its performance across different datasets and network topologies, including varying demand matrices, interference conditions, and mobility patterns representative of realistic B5G environments. The model will be further enhanced to improve adaptability and generalization across both short and long-term traffic prediction tasks under dynamic network conditions. Additionally, we plan to incorporate reinforcement learning–based policy adaptation and automated hyperparameter optimization to enhance the framework’s scalability, stability, and end-to-end decision predicting accuracy.

\section*{Acknowledgment}
This work is supported by the ZU ViP project EU2105 and the ZU Provost’s Research Fellowship Award (PRFA) 23061.

%In this paper, we discuss different Generative AI approaches available for Networking. We also highlighted different areas that need to be investigated for the use of GenAi. Finally, we provide a use case for a transformer-based long-term network prediction model. This model leverages a self-attention mechanism to correlate different input parameters to determine the output sequence. The results show that long-term prediction is better than current state-of-the-art such as transformers and informers. In the future, we will include the policy generation mechanism and different network slice parameters to improve the accuracy of the model. We will also integrate the solution with an emulated test bed using Mininet. 

\end{document}